\documentclass[fleqn,usenatbib]{mnras}

\usepackage{newtxtext,newtxmath}

\usepackage[T1]{fontenc}

\DeclareRobustCommand{\VAN}[3]{#2}
\let\VANthebibliography\thebibliography
\def\thebibliography{\DeclareRobustCommand{\VAN}[3]{##3}\VANthebibliography}

\def\cms{\hbox{\rm\hskip.35em  cm s}$^{-1}$}

\usepackage{graphicx}	
\usepackage{amsmath}	
\usepackage{threeparttable}

\title[Rotation of PSR B0950+08]{On the Peculiar Rotational Evolution of PSR B0950+08}

\author[G\"{u}gercino\u{g}lu et al.]{
Erbil G\"{u}gercino\u{g}lu,$^{1,2}$\thanks{E-mail: egugercinoglu@gmail.com (EG)}
Elif K\"{o}ksal$^{3}$
and Tolga G\"{u}ver$^{2,4}$
\\
$^{1}$National Astronomical Observatories, Chinese Academy of Sciences, 20A Datun Road, Chaoyang District, Beijing 100101, China\\
$^{2}$Istanbul University, Faculty of Science, Department of Astronomy and Space Sciences, Beyaz{\i}t, 34119, Istanbul, Turkey \\
$^{3}$Istanbul University, Institute of Graduate Studies in Science, Programme of Astronomy and Space Sciences, 34116, Beyaz{\i}t, Istanbul, Turkey\\
$^{4}$Istanbul University Observatory Research and Application Center, Istanbul University 34119, Istanbul Turkey
}

\date{Accepted 2022 November 29. Received 2022 November 28; in original form 2022 July 7}

\pubyear{2023}

\begin{document}
\label{firstpage}
\pagerange{\pageref{firstpage}--\pageref{lastpage}}
\maketitle

\begin{abstract}
The long-term rotational evolution of the old, isolated pulsar, PSR B0950+08 is intriguing in that its spin-down rate displays sinusoidal-like oscillations due to alternating variations, both in magnitude and sign, of the second time derivative of the pulse frequency. We show that the large internal temperature to pinning energy ratio towards the base of the crust implied by the recent high surface temperature measurement of PSR B0950+08 leads to linear creep interaction between vortex lines and pinning sites to operate in this pulsar. Vortex lines assume a parabolic shape due to pinning to nuclear clusters and finite tension of vortices acts as a restoring force that tends to bring a vortex back to its straight shape. The resulting low frequency oscillations of vortex lines combined with the time variable coupling between the internal superfluid components and the external pulsar braking torque give rise to an oscillatory spin-down rate. We apply this model to PSR B0950+08 observations for several external torque models. Our model has potential to constrain the radial extension of the closed magnetic field region in the outer core of neutron stars from the oscillation period of the spin-down rate. 
\end{abstract}

\begin{keywords}
stars: neutron -- pulsars: general -- pulsars: individual: PSR B0950+08
\end{keywords}



\section{Introduction}

The long-term timing observations of pulsars provide important information on the magnetospheric emission mechanisms \citep{wu03,kou15} and internal superfluid dynamics \citep{alpar06,lower21} of neutron stars. A key parameter of pulsar spin-down is the braking index $n$, which is defined in terms of its pulse frequency $\nu$, spin-down-rate $\dot\nu$ and second time derivative of frequency $\ddot\nu$ as \citep{blandford88}
\begin{equation}
n=\frac{\nu\ddot\nu}{\dot\nu^{2}}.
\label{nconstant}
\end{equation} 
Among the spin parameters, an unambiguous measurement of $\ddot\nu$ requires long-term and high cadence monitoring of a given pulsar. 

PSR B0950+08 (J0953+0755) is an isolated radio pulsar discovered by \citet{pilkington68} with spin frequency $\nu=3.95$~Hz, spin-down rate $\dot\nu=-3.59\times10^{-15}$~Hz\,s$^{-1}$, characteristic (spin-down) age $\tau_{\rm sd}=\nu/(2|\dot\nu|)=17.5$~ Myr, and inferred surface dipole magnetic field of $B_{\rm s}=2 .44\times10^{11}$ G. PSR B0950+08 is a very peculiar source among old isolated radio pulsars in many respects. 

There is considerable uncertainty regarding the true age of PSR B0950+08. \citet{noutsos13} estimated by examining spin-space velocity alignment that its kinematic age was smaller than the characteristic age quoted above. Using a Bayesian approach and considering putative magnetic field decay and long initial spin period, \citet{igoshev19} arrived at the conclusion that the most likely kinematic age of PSR B0950+08 might be $1.9^{+5.5}_{-0.6}$~Myr with 68 per cent (1-$\sigma$) confidence level. This value is again significantly lower than the characteristic age of PSR B0950+08. Thus, kinematic age considerations with possible magnetic field decay indicate that PSR B0950+08 may be younger than its spin-down age implies.

\citet{abramkin22} analysed far-ultraviolet and optical spectrum of PSR B0950+08 with blackbody thermal component and nonthermal power law fits and placed an upper limit $T_{\infty}<1.7\times10^{5}$ K for its surface temperature as seen by a distant observer. For standard cooling scenarios \citep{yakovlev04,page06,yanagi20}, neutron star surface temperatures are expected to fall well below $10^{4}$~K for ages $\gtrsim1$~Myr unless some heating mechanisms operate inside them. In old neutron stars, superfluid friction with normal matter and rotochemical heating are the two main dissipative processes generating higher temperatures \citep{alpar84a,reisenegger10}.

Recently, \citet{huang22} have presented a timing solution of PSR B0950+08, using 14 yr of observations from the Nanshan 26-m Radio Telescope of Xinjiang Astronomical Observatory. Most notably, its spin-down rate exhibits sinusoidal-like oscillation (see figures 1 and 2 of \citet{huang22} and Fig. \ref{fig:fit} below). As a consequence of oscillatory spin-down, the braking index of PSR B0950+08 was found to alternate between $-367 392$ to $168 883$ with a large amplitude of variation \citep{huang22}. Since in the same time span no glitch was reported, the observed oscillation in the spin-down rate should reflect the steady dynamical behaviour of this old pulsar and can be used to investigate various features of neutron star. Below, we utilize the spin-down rate data points given in table 2 of \citet{huang22} to apply our model, which is developed for the long-term spin evolution of pulsars.

Unmodelled timing noise from middle-aged and old pulsars may lead to slow, discernable stochastic wandering of spin parameters at different levels \citep{arzoumanian94,hobbs10,lower20}. For PSR B0950+08, \citet{shaw22} have recently reported on a noise variance of $\sigma^{2}_{\rm N}=2.7\times10^{-9}$ s after analysing about 40 yr of Jodrell Bank data. However, the stability of timing residuals, i.e. the absence of a clear trend in the timing solution after removal of an assumed model [see figure 1 in \citet{shaw22}], implies that there is no appreciable white noise or random walk in the torque acting on PSR B0950+08 but a second order red noise in the torque may be present if at all \citep{baykal99}. From figure 1 of \citet{shaw22}, it appears that the deviation of phase residuals from the assumed timing model for PSR B0950+08 is modest for the range from MJD 51547 to MJD 56664 compared to the amplitude of the residuals of this pulsar at other epochs. The same time span of data was also analysed in \citet{huang22} and the corresponding data were considered in this study. We keep in mind that some level of noise may be present in the long-term spin evolution of PSR B0950+08.

\citet{huang22} studied the coupled spin and thermal evolution of PSR B0950+08  by elaborating on the combined effects of magnetic field decay and vortex creep heating. Vortex creep is thermally activated motion of vortex lines against potential barriers sustained by the lattice nuclei under the bias of secular neutron star spin-down \citep{alpar84a}. Vortex creep depletes differential rotation between the superfluid and the normal matter, thereby dissipates the rotational energy of the neutron star and heats it up. \citet{huang22} consider the long-term magnetic field decay, which is modulated by short-term oscillations as the main cause of braking index variation, and take heating due to magnetic field decay and vortex creep  into account in order to explain high surface temperature measurement. \citet{zhang12} proposed that phenomenological short-term oscillatory behaviour of the magnetic field superimposed on long-term decay may be the underlying reason for the large discrepancies among measured braking indices of middle aged and old isolated pulsars.  

In this study, we consider the peculiar rotational evolution of PSR B0950+08 by invoking time variable interior superfluid coupling to the pulsar braking torque. In Section \ref{sec:model}, we outline model equations. In Section \ref{sec:results}, we apply our model equations to the long-term spin-down rate data of PSR B0950+08. In Section \ref{sec:conclusions}, we discuss our conclusions. 

\section{Model Equations}
\label{sec:model}

\begin{figure}	
	\includegraphics[width=\columnwidth]{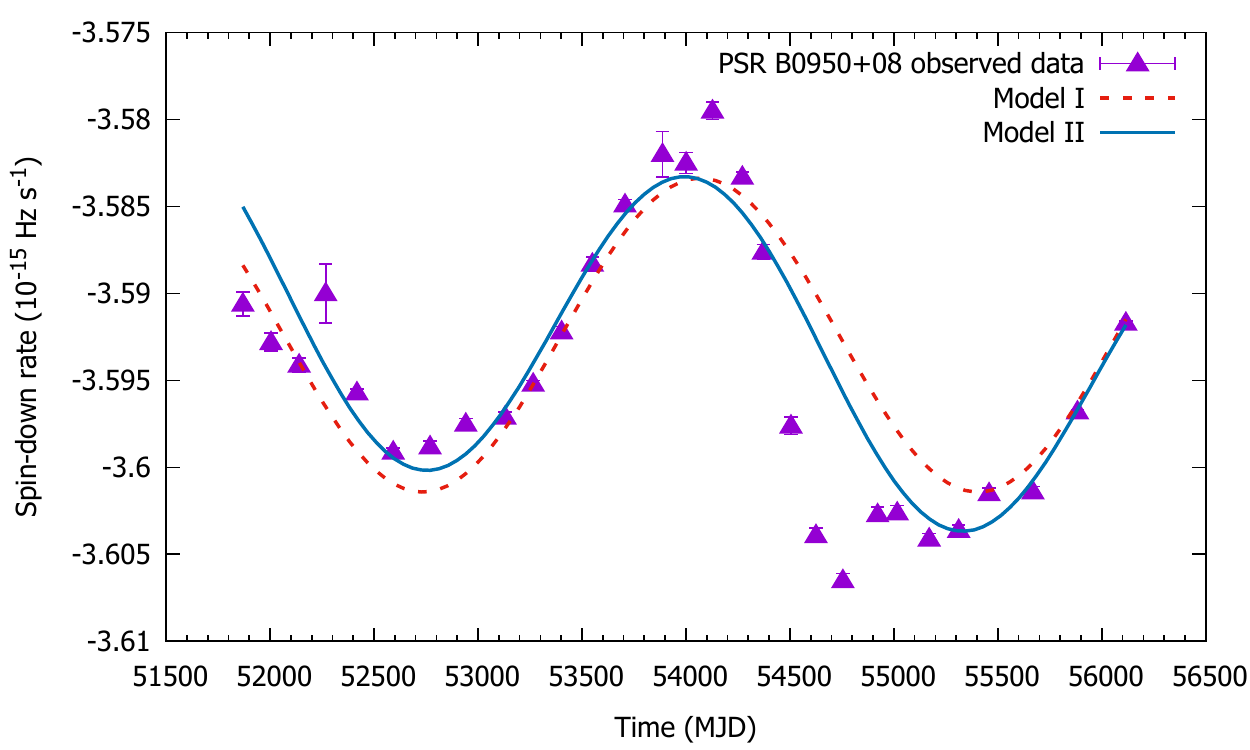}
    \caption{The long-term spin-down rate evolution of PSR B0950+08 and the vortex creep model fits to the data. Observational data points with error bars (purple triangles) are taken from table 2 of \citet{huang22}. Model I (dashed red curve) is for constant external torque, while Model II (solid blue curve) corresponds to the negative braking index case. See Section \ref{sec:model} for details regarding the model equations.}
    \label{fig:fit}
\end{figure}

\begin{figure}
	\includegraphics[width=\columnwidth]{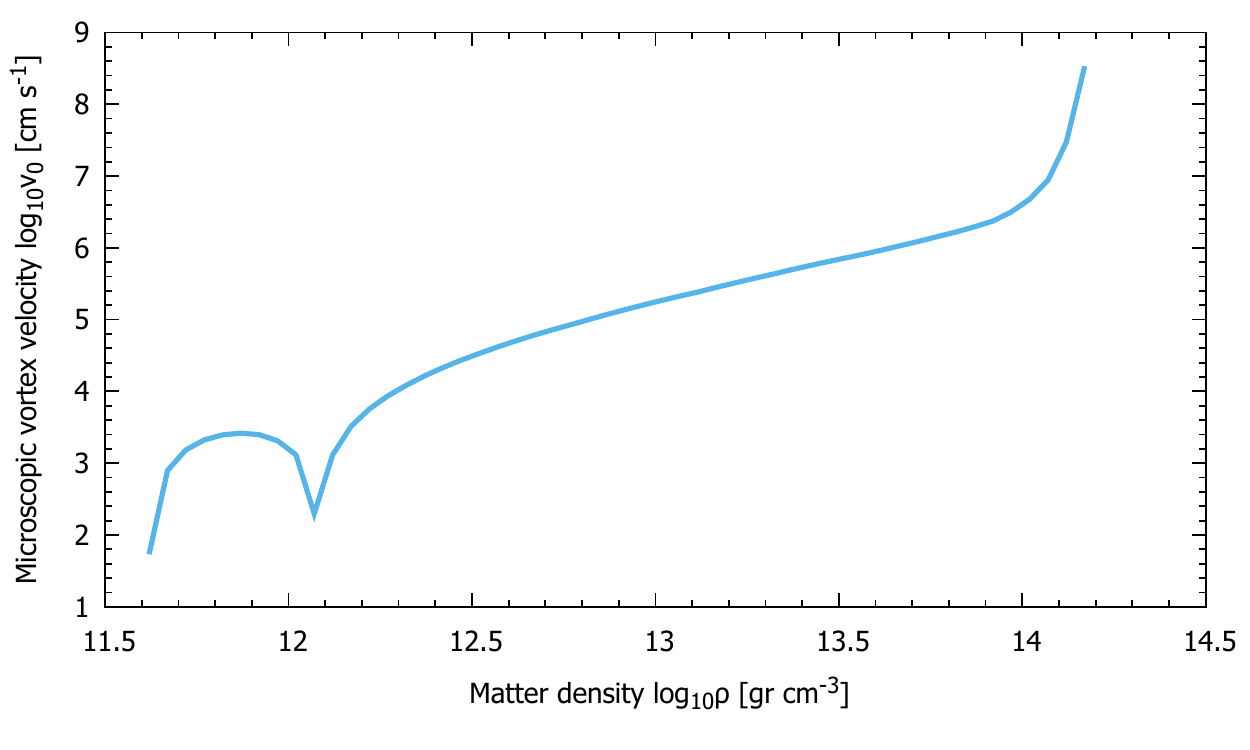}
    \caption{Microscopic vortex velocity $v_{0}$ in the crustal superfluid versus matter density $\rho$. Equation of state related microphysical parameters are taken from \citet{kobyakov16}. See \citet{erbil16} for details.}
    \label{fig:v0}
\end{figure}

The rotational evolution equations describing a three-component neutron star consisting of normal matter (subscript “n”), crustal superfluid (subscript “cs”) and core superfluid (subscript “core”) can be expressed as \citep{lamb78,erbil17}
\begin{equation}
I_{\rm n}\dot\Omega_{\rm n}=N_{\rm ext}(t)-I_{\rm cs}\dot\Omega_{\rm cs}-I_{\rm core}\dot\Omega_{\rm core}, 
\label{evolutionn}
\end{equation}
\begin{equation}
\dot\Omega_{\rm cs}=-\frac{\varpi}{\tau_{\rm l}}\sinh\left(\frac{\Omega_{\rm cs}-\Omega_{\rm n}}{\varpi}\right)-\frac{\vartheta_{0}^{2}}{I_{\rm cs}}\int^{t}dt'\left(\Omega_{\rm cs}-\Omega_{\rm core}\right),
\label{evolutioncs}
\end{equation}
\begin{equation}
\dot\Omega_{\rm core}=-\frac{\left(\Omega_{\rm core}-\Omega_{\rm n}\right)}{\tau_{\rm core}}+\frac{\vartheta_{0}^{2}}{I_{\rm core}}\int^{t}dt'\left(\Omega_{\rm cs}-\Omega_{\rm core}\right),
\label{evolutioncore}
\end{equation}
where $I_{\rm n}, I_{\rm cs}$, and $I_{\rm core}$ are moments of inertia of the corresponding components with $\Omega_{\rm n}, \Omega_{\rm cs}$, and $\Omega_{\rm core}$ being their respective angular rotation rates, respectively. 

Equation (\ref{evolutionn}) describes the torque equilibrium on a neutron star under external magnetospheric braking torque $N_{\rm ext}$ and internal torques due to crustal and core superfluids. In equation (\ref{evolutioncs}), $\varpi\equiv(kT/ E_{\rm p})\omega_{\rm cr}$, where $k$ is the Boltzmann constant, $T$ is the internal temperature, $E_{\rm p}$ is the pinning energy, and $\omega_{\rm cr}$ is the critical angular velocity between the rotation rates of the crustal superfluid and the normal matter that vortex lines can withstand before unpinning. Linear creep time-scale $\tau_{\rm l}$ is expressed in terms of crustal microphysical parameters as \citep{alpar89}
\begin{equation}
\tau_{\rm l}=\left(\frac{kT}{E_{\rm p}}\right)\frac{\omega_{\rm cr}r}{4\Omega_{\rm cs}v_{0}}\exp\left(\frac{E_{\rm p}}{kT}\right),
\label{taulin}
\end{equation} 
where $r$ is the distance of vortex lines from the rotational axis and $v_{0}$ is the microscopic vortex velocity around the nuclear clusters. The variation of the microscopic velocity with the crustal density is shown in Fig. \ref{fig:v0}.   

In equation (\ref{evolutioncore}), $\tau_{\rm core}$ is the time-scale for communicating the changes in the crustal rotation rate to the core superfluid. In the core superfluid, proton entrainment mass currents circulate around a vortex line. Such circulation endows each vortex line with a huge quantized magnetic field of the order of $10^{15}$\,G, which decays over a length-scale determined by the London penetration depth $\Lambda\sim10^{-11}$\,cm of the type II proton superconductor inside the neutron star core. The scattering of electrons from the magnetized vortices maintains an efficient coupling between the crustal normal matter and the core superfluid \citep{alpar84b}. The prompt relaxation of the rotation rate immediately after the 2016 Vela glitch \citep{ashton19} implies a time variable fast coupling of the core superfluid to the observed crust \citep{graber18,pizzochero20,erbil20}.  

In equations (\ref{evolutioncs}) and (\ref{evolutioncore}), $\vartheta_{0}$ is related to the oscillation mode of the internal superfluid due to bending of vortex lines. A vortex line crosses the whole neutron star and points in the direction of the rotation axis in order to carry circulation within the superfluid, and an array of vortices collectively maintains global rotation. While part of a given vortex lying in the neutron star inner core is straight, the remaining length of the same line is slightly bent in the inner crust and outer core regions due to pinning to lattice nuclei \citep{alpar77,alpar84a} and magnetic flux tubes \citep{sidery09,erbil14}, respectively. Vortex rigidity due to self-tension acts as restoring force compensating for the bending of the vortex that occurs as a result of the pinning interaction. A simple argument leads to \citep{sedrakian95}\footnote{Here, we have corrected a typo in the expression for vortex self-energy per unit length used in \citet{sedrakian95}, which erroneously leads to a factor of $2\pi$ larger vortex oscillation frequencies $\sqrt{\vartheta_{0}^{2}/I_{\rm s}}$.}
\begin{equation}
\vartheta_{0}^{2}\simeq\frac{10\Omega_{\rm n}\kappa I_{\rm s}}{\pi\ell^{2}},
\label{bentvor}
\end{equation} 
where $\kappa=2\times10^{-3}$\,cm$^{2}$\,s$^{-1}$\, is the quantized vorticity attached to the each line and $\ell$ is the length of the part of the vortex that is bent. The part of the moment of inertia $ I_{\rm s}$ of the superfluid components in which vortex lines experience bending due to pinning, i.e. the crustal superfluid plus the outer core superfluid, can be determined from the evaluation of the following integral over the corresponding portion of superfluids
\begin{equation} 
 I_{\rm s}=\int_{r_{0}}^{R_{0}}\rho_{\rm s}r^{2}\ell2\pi r dr,
\end{equation}
where $\rho_{\rm s}$ is the superfluid mass density, $r_{0}$ is the radius of the bottom edge of the outer core (closed magnetic field region) harbouring toroidal flux tubes, and $R_{0}$ is the outer radius of the crustal superfluid in the linear regime.

The vortex length permeating the inner crust superfluid can be expressed in terms of crustal thickness $\Delta R$ and neutron star radius $R$ as \citep{link02}
\begin{equation}
\ell_{\rm crust}=\frac{\Delta R}{2}+R\sqrt{\frac{\Delta R}{2R}}.
\end{equation}
For a typical neutron star, $\ell_{\rm crust}$ is about few times $10^{5}$ cm.

Depending on the linearity parameter $\eta$
\begin{equation}
\eta=\frac{|\dot\Omega_{\infty}|r}{4\Omega_{\rm cs}v_{0}}\exp\left(\frac{E_{\rm p}}{kT}\right)\equiv\frac{\tau_{\rm l}}{\tau_{\rm nl}},
\end{equation}
 vortex creep is either in the linear ($\eta<1$) or non-linear ($\eta>1$) regime. Here, $\dot\Omega_{\infty}$ is the steady-state spin-down rate, $\tau_{\rm nl}\equiv\varpi/|\dot\Omega_{\infty}|$ is the non-linear superfluid recoupling time-scale, and linear creep time-scale $\tau_{\rm l}$ is given by equation (\ref{taulin}). The linear to non-linear creep transition occurs at \citep{alpar89}
\begin{equation}
\left(\frac{E_{\rm p}}{kT}\right)_{\rm tr}=\ln\left(8\tau_{\rm sd}\frac{v_{0}}{r}\right)=35.46+\ln\tau_{\rm sd,6}+\ln\left(\frac{v_{0,7}}{r_{6}}\right),
\label{lnlcreep}
\end{equation}
where $\tau_{\rm sd,6}$ is the characteristic age $\tau_{\rm sd}=\Omega_{\rm n}/(2|\dot\Omega_{\rm n}|)$ in units of $10^{6}$ yr, $v_{0,7}$ is the microscopic vortex velocity in units of $10^{7}$\cms,~and $r_{6}$ is the vortex line distance from the rotation axis in units of $10^{6}$ cm. To determine the vortex creep regime for PSR B0950+08, one needs to compare effective pinning energy with internal crustal temperature. Recently, \citet{abramkin22} obtained the range $6<T_{\rm s}/10^{4} \mbox{K}<16$ for the surface temperature of PSR B0950+08 by combining the latest far-ultraviolet and optical spectral observations. The surface temperature may be converted to internal temperature by the following formula, which is valid for a non-magnetic Fe atmosphere neutron star model \citep{gudmundsson82}
\begin{equation}
T_{8}=1.288\left(\frac{T_{\rm s6}^{4}}{g_{\rm s14}}\right)^{0.455}.
\label{Tint}
\end{equation}
Here, we use the standard notation that $Q_{\rm x} = Q/10^{{\rm x}}$ for the corresponding quantity $Q$ in cgs units. For a $1.4M_{\odot}$ neutron star, the SLy4  equation of state {yields a gravitational redshift-corrected surface gravity (in units of $10^{14}$\,cm\,s$^{-2}$) of $g_{\rm s14}=1.78$ \citep{haensel01}. Then, the surface temperature constraint obtained by \citet{abramkin22} translates into internal temperature range $0.6<T/10^{6} \mbox{K}<3.7$ for PSR B0950+08 by equation (\ref{Tint}). With $\tau_{\rm sd,6}=17.5$ and $v_{0,7}$ values from Fig. \ref{fig:v0}, equation (\ref{lnlcreep}) yields the transition effective pinning energy bound $E_{\rm eff,tr}\lesssim13.2$ keV for PSR B0950+08. When the effective pinning energy in a particular crustal superfluid layer is less than this transition value, the spin-down of the corresponding superfluid layer linearly depends on the lag between the observed crustal and superfluid layer's rotation rates in question.

As initial calculations indicate, the pinning energy between a vortex line and lattice nuclei decreases with increasing density \citep{alpar77}. Towards the base of the crust, the neutron pairing gap reduces due to proximity effects on the pairing correlations \citep{gandolfi08,urban20,matsuo21}. The equilibrium configuration and shape of a vortex under pinning forces is determined from the Magnus force law with proper inclusion of the effects of vortex self-energy \citep{link91,shibazaki01}. The finite tension $T_{\rm v}\approx\rho_{\rm s}\kappa^{2}/4\pi$ of a vortex sets the length-scale over which part of the vortex can interact with nuclei as it determines the bending of the line. When the effect of vortex tension is taken into account, the effective pinning energy scales as $E_{\rm p}\propto T_{\rm v}^{-1/2}$ \citep{link91}. An assessment of correlated creep rate and equilibrium vortex line configuration for the densest pinning layers in the crust shows that vortices assume a parabolic shape in the neighbourhood of lattice nuclei, which prevents large pinning energies \citep{chau93}. When all these effects are considered together, the effective pinning energy is of the order of $E_{\rm p}\sim10$~keV for densities $\rho\sim10^{14}$\,g\,cm$^{-3}$ where most of the moment of inertia of the neutron star crustal superfluid resides \citep{seveso16}. \citet{shaw22} measured the peak-to-peak fractional amplitude of the variations in the spin-down rate of PSR B0950+08 as $\Delta\dot\nu/\dot\nu=0.8$ per cent, which should be equal to the fractional moment of inertia of the loosely coupled superfluid component that plays a role in the torque oscillations. This is again in line with our estimate of linear crustal superfluid amount in the range $14.0\lesssim\log_{10}\rho (\mbox{g cm}^{-3})\lesssim14.2$. When the crustal entrainment [see e.g. \cite{chamel17} and references therein] is taken into account, the range can be extended to slightly lower densities as this effect will somewhat reduce the mobility of the unbound neutrons. Therefore, for heated old pulsars like PSR B0950+08, vortex creep is expected to be in the linear regime in the densest pinning layers of the crust. 

In the limit of strong coupling between the normal matter and the core superfluid (i.e. $\tau_{\rm core}\rightarrow0$) and if $\tau_{\rm l}<\tau_{\rm nl}\equiv\varpi/|\dot\Omega_{\infty}|$, equations (\ref{evolutionn})-(\ref{evolutioncore}) reduce to the following set of equations:
\begin{equation}
I_{\rm c}\dot\Omega_{\rm c}=N_{\rm ext}(t)-I_{\rm cs}\dot\Omega_{\rm cs}, 
\label{limeq1}
\end{equation}
\begin{equation}
\dot\Omega_{\rm cs}=-\frac{\left(\Omega_{\rm cs}-\Omega_{\rm c}\right)}{\tau_{\rm l}}-\frac{\vartheta_{0}^{2}}{I_{\rm cs}}\int^{t}dt'\left(\Omega_{\rm cs}-\Omega_{\rm c}\right),
\label{limeq2}
\end{equation}
where $I_{\rm c}=I_{\rm n}+I_{\rm core}$ with $I=I_{\rm c}+I_{\rm cs}$ being the total moment of inertia of the neutron star and $\Omega_{\rm c}$ is the common rotation rate of the observed crust and core superfluid.
The last two equations, namely (\ref{limeq1}) and (\ref{limeq2}), can be combined to yield the following second-order differential equation for the rotation rate of the observed crust:
\begin{equation}
\ddot\Omega_{\rm c}+\frac{\dot\Omega_{\rm c}}{\tau}+\omega_{0}^{2}\Omega_{\rm c}=\frac{\dot N_{\rm ext}(t)}{I_{\rm c}}+\frac{N_{\rm ext}(t)}{I\tau}+\frac{\omega_{0}^{2}}{I}\int^{t}dt'N_{\rm ext}(t'),
\label{modeleq}
\end{equation}
where we have defined
\begin{equation}
\omega_{0}^{2}\equiv\frac{I}{I_{\rm c}I_{\rm cs}}\vartheta_{0}^{2}.
\label{oscfre}
\end{equation}
Here, $I/I_{\rm c}\cong1$. Note that equation (\ref{modeleq}) is also obtained in \citet{lamb78} but with a phenomenological form of $\tau$. In our case, $\tau=(I_{\rm c}/I)\tau_{\rm l}\cong\tau_{\rm l}$ is the linear regime recoupling time-scale of the crustal superfluid that has direct relation with microphysical properties and superfluid traits of the neutron star crust. 

In the next two subsections, we consider two forms of $N_{\rm ext}(t)$: constant external torque and time variable braking due to magnetic field change.

\begin{table*}
\caption{Fit parameters corresponding to Model I [equation (\ref{model1})] and Model II [equation (\ref{model2})].}
\begin{center}
\begin{threeparttable}
\begin{tabular}{lcc}

\hline
\hline  
Parameter                                                         & Vortex Creep Model I               & Vortex Creep Model II                 \\
\hline
$A$ (10$^{-18}$\,rad\,s$^{-1}$)                    &    1.75$\pm$0.29                     &     1.74$\pm$0.23                         \\
$\tau (10^{8}$ days)                                       &    9.19                                      &     6.65                                             \\
$\Omega_{0}$ ($10^{-8}$ rad\,s$^{-1}$)        &    2.72$\pm$0.05                           &     2.82$\pm$0.04                        \\
$\phi$ (rad)                                                        &    1.11$\pm$0.14                     &     1.05$\pm$0.10                                    \\
$\dot\nu_{0}$ (10$^{-15}$\,Hz\,s$^{-1}$)   &    $-3.5924\pm$0.0006             &     $-3.5897\pm$0.0009                     \\
$\ddot\nu_{0}$ (10$^{-26}$\,Hz\,s$^{-2}$) &       -                                          &    $-1.57\pm$0.41                            \\

\hline

\end{tabular}
\end{threeparttable}
\end{center}
\label{tab:fit}
\end{table*}

\subsection{Model I: Constant External Torque}
\label{sec:torque1} 

Given that the external braking torque acting upon pulsars changes slowly on the spin-down time-scale, one can assume $N_{\rm ext}(t)=2\pi I\dot\nu_{0}$ to be constant on much shorter 14 yr of observation for old PSR B0950+08. For this choice, the solution of equation (\ref{modeleq}) becomes
\begin{equation}
\Omega_{\rm c}=A\exp\left(-\frac{t}{2\tau}\right)\sin\left(\Omega_{0} t+\phi\right)+2\pi\dot\nu_{0}t,
\label{eqconstant}
\end{equation}
which is in the same mathematical form as a damped harmonic oscillator [c.f. section 5 of \citet{landau69}] modulated around the value $2\pi\dot\nu_{0}t$. Here, the amplitude $A$ and the phase $\phi$ are integration constants. We have also introduced the oscillation frequency of the rotation of the neutron star associated with the coupling of internal superfluid modes and normal matter crust
\begin{equation}
\Omega_{0}=\omega_{0}\left[1-\left(\frac{1}{2\tau\omega_{0}}\right)^{2}\right]^{1/2}.
\label{Oscfre}
\end{equation}
The time derivative of equation (\ref{eqconstant}) gives spin-down rate, which can be directly compared with observations:
\begin{align}
\dot\Omega_{\rm c}(t)=&A\Omega_{0}\exp\left(-\frac{t}{2\tau}\right)\cos\left(\Omega_{0} t+\phi\right)-\frac{A}{2\tau}\exp\left(-\frac{t}{2\tau}\right)\sin\left(\Omega_{0} t+\phi\right) \nonumber\\&+2\pi\dot\nu_{0}.
\label{model1}
\end{align}
This expression is formally identical to the solution obtained in \citet{sedrakian95} for modelling the post-glitch fluctuations in the rotation and spin-down rates observed after the Christmas 1988 Vela glitch.
\subsection{Model II: Negative Braking Index Due to Magnetic Field Change}

Magnetic fields of neutron stars decay in time due to processes prevailing in their crusts and cores \citep{goldreich92,pons19,igoshev21}. For isolated canonical neutron stars, Ohmic diffusion dissipates the crustal magnetic field while interpinning of vortex lines to the magnetic flux tubes expels the corresponding flux out of the core \citep{ding93}. The flux entry into the crust occurs at about Ohmic diffusion time-scale $\tau_{\rm Ohm}$, which is estimated to be a few Myr \citep{igoshev21}. Therefore, in old pulsars, field growth may take place superimposed on the long-term field decay.

If the magnetic field of a pulsar changes with time, its braking index becomes \citep{pons12}
\begin{equation}
n=3-4\frac{\dot B}{B}\tau_{\rm sd},
\label{nbchange}
\end{equation}
where $\dot B$ is the time derivative of the magnetic field $B$. Note that $n>3$ means that neutron star magnetic field decays in time, while $n<3$ requires magnetic field growth. The simplistic choice of the form of $N_{\rm ext}$ that includes the effects of change of the magnetic field strength in accordance with equations (\ref{nconstant}) and (\ref{nbchange}) is
\begin{equation}
N_{\rm ext}=2\pi I(\dot\nu_{0}+\ddot\nu_{0}t),
\label{NextB}
\end{equation}
where $\ddot\nu_{0}$ encodes the effects of the magnetic field change. With equation (\ref{NextB}) the solution of equation (\ref{modeleq}) gives
\begin{equation}
\Omega_{\rm c}=A\exp\left(-\frac{t}{2\tau}\right)\sin\left(\Omega_{0} t+\phi\right)+2\pi t\left(\dot\nu_{0}+\frac{1}{2}\ddot\nu_{0}t\right)
\end{equation}
and the measured spin-down rate becomes
\begin{align}
\dot\Omega_{\rm c}(t)=&A\Omega_{0}\exp\left(-\frac{t}{2\tau}\right)\cos\left(\Omega_{0} t+\phi\right)-\frac{A}{2\tau}\exp\left(-\frac{t}{2\tau}\right)\sin\left(\Omega_{0} t+\phi\right) \nonumber\\&+2\pi\dot\nu_{0}+2\pi\ddot\nu_{0}t.
\label{model2}
\end{align}

\section{Results}
\label{sec:results}

We apply the equations, namely (\ref{model1}) and (\ref{model2}) corresponding to Model I and Model II, obtained in the previous section to the spin-down rate behaviour of PSR B0950+08 by taking $\dot\nu(t)=\dot\Omega_{\rm c}(t)/2\pi$. Model fits to the observed spin-down rate data, dashed red curve for Model I and solid blue curve for Model II, are shown in Fig. \ref{fig:fit}. The fit parameters are given in Table \ref{tab:fit}. While the parameters do not differ appreciably, Model II provides better fit to the observations due to inclusion of the efficacy of the magnetic field change on the second time derivative of the pulse frequency $\ddot\nu_{0}$. This is reflected in the reduced $\chi^{2}$ values: 8.2 for Model I, whereas 5.7 for Model II.

For the constant external torque (Model I) case, the spin-down rate displays damped oscillations around the value $\dot\nu_{0}=-3.5924\times10^{-15}$\,Hz\,s$^{-1}$.  Since only one cycle is completed in Fig. \ref{fig:fit}, the damping time-scale cannot be identified precisely. For the Jodrell Bank observations of PSR B0950+08, three peaks in the spin-down rate were reported for a longer data span that occurred around MJD 44000, MJD 48000, MJD 54000 but no consecutive oscillations were detected \citep{shaw22}\footnote{Note, however, that towards the end of data set of \citet{shaw22} there also exists a clear increase in the spin-down rate beginning around MJD 56000, which appears to have somewhat different form from the previous three oscillations. The recent charge reconfiguration in the force-free magnetosphere and its coupling with the various internal superfluid components or the effects of different origin may lead to clear deviation of the spin-down rate from its steady value. Future observations will help to understand the true underlying behaviour.}. This fact puts a limit on the damping time-scale of oscillations. The length-scale for the bent vortex segment is determined from Table \ref{tab:fit} and equations (\ref{bentvor}), (\ref{oscfre}), and (\ref{Oscfre}) as $\ell=4\times10^{5}$~cm. Therefore, straight vortex lines should be bent over nuclear pinning region in the inner crust and toroidal flux tube region in the outer core. $T_{\rm osc}=2\pi/\Omega_{0}$ gives the oscillation period in the spin-down rate due to combined effects of internal superfluid modes associated with vortex bending and its coupling with external braking torque. Collective vortex lattice vibrations, known as Tkachenko modes, lead to long period oscillations for which the fundamental mode is given by \citep{ruderman70} 
\begin{equation}
T_{\rm Tkachenko}=\frac{140R}{\Omega^{1/2}} ~\mbox{s},
\label{Posc}
\end{equation}
where $R$ is the neutron star radius and $\Omega$ is the angular rotational velocity.  From Table \ref{tab:fit}, $T_{\rm osc}=7.06$~yr, while equation (\ref{Posc}) gives $T_{\rm Tkachenko}=0.89$\,yr for PSR B0950+08 with $\Omega=2\pi\nu\cong25$\,rad\,s$^{-1}$ and $R=10^{6}$\,cm. Therefore, the low frequency vortex oscillations leading to sinusoidal-like spin-down rate behaviour for PSR B0950+08 cannot solely be due to fundamental Tkachenko mode and some mixed overtones should be involved. 
Another vortex oscillation mode is associated with the spherical geometry of the crust. Since vortex lines have cylindrical geometry, the vortices approaching to the neutron star equator region should be slightly curved and get shorter. The resulting vortex mode has an Ekman oscillation period given by \citep{alpar78}
\begin{equation}
T_{\rm Ekman}=\frac{R}{\left(\kappa\Omega\right)^{1/2}}.
\label{Pekman}
\end{equation}
Equation (\ref{Pekman}) gives $T_{\rm Ekman}=52$ d for PSR B0950+08 and has no relevance for 14 yr period of observations. 

The other superfluidity related modes, namely collective vortex oscillations, i.e. Tkachenko modes, and Ekman pumping lead to $\sim1$\,yr and unobservationally short 52 d time-scales, respectively, for the oscillation period, both of which are in sharp conflict with the observations of PSR B0950+08. This conclusion is also valid for the remaining pulsars in the sample of \citet{shaw22} (K\"{o}ksal et al., in preparation). Also, adjusting the deviation of magnetospheric charge density from the Goldreich-Julian value \citep{kramer06} and various decay modes superimposed on the long-term magnetic field evolution of neutron stars \citep{biryukov12} require some level of fine-tuning. Therefore, we can safely argue that the vortex bending due to the presence of nuclear clusters in the crust and toroidal arrangement of flux tubes in the core gives rise to dynamically important consequences for neutron star spin evolution. As a pulsar ages, the contributions from the external braking torque and superfluid torque with vortex bending assisted oscillation modes to the pulsar spin-down become comparable in magnitude. This may account for the anomalous (large and/or negative) braking indices and fluctuations in the second time derivative of spin frequency $\ddot\nu$ seen across the pulsar population \citep{hobbs10,parthasarathy19}.

For Model II, $\ddot\nu_{0}=-1.57\times10^{-26}$\,Hz\,s$^{-2}$ and $\dot\nu_{0}=-3.59\times10^{-15}$\,Hz\,s$^{-1}$ imply a braking index of $n=-4815$, which in turn gives $\tau_{\rm B}=B/\dot B\cong1.45\times10^{2}$~yr for the magnetic field growth time-scale via equation (\ref{nbchange}). Thus, a recent transportation of the some of the core magnetic flux into the crust by the secular radially outward motion of superfluid vortex lines may lead to negative $\ddot\nu$.

There are some correlations among fit parameters. For Model I, the oscillation frequency of internal superfluid due to vortex bending $\Omega_{0}$ shows a positive correlation with the spin down rate, which corresponds to the offset in the sinusoidal change. For Model II, a strong negative correlation exists between $\dot\nu_{0}$ and $\ddot\nu_{0}$. Thus, it can be inferred that if the effect of magnetic field's change is increased, the spin-down rate will decrease. This is exactly as expected since in our model equations, $\ddot\nu_{0}/\dot\nu_{0}\sim1/\tau_{\rm B}$. Moreover,  the oscillation frequency of internal superfluid due to vortex bending $\Omega_{0}$ shows a negative relation with the spin-down rate $\dot\nu_{0}$. Also, there is a positive relation between the oscillation amplitude $A$ and the second time derivative of the pulse frequency $\ddot\nu_{0}$ for Model II. This can be easily understood because $\ddot\nu_{0}$ tends to track the data points downward, while the amplitude $A$ compensates for the increase trend of the spin-down rate.

\section{Conclusions}
\label{sec:conclusions}

Superfluid vortex creep heating in old pulsars with characteristic spin-down-ages $\tau_{\rm sd}\gtrsim1$ Myr leads to a change in the response of the interior torque for the densest pinning layers of the inner crust superfluid. Combination of the linear creep response with low frequency vortex line oscillation modes due to bending of lines in the pinning regions of inner crust and outer core superfluids results in damped sinusoidal-like oscillations in the spin-down rate of pulsars. Inclusion of the pulsar braking torque determines the time variable external magnetospheric and interior superfluid torques coupling which in turn allows for identification of neutron star equation of state-related physical parameters and processes prevailing in the magnetosphere.

The sinusoidal oscillation period $T_{\rm osc}=2\pi/\Omega_{0}$ of the spin-down rate of pulsars with $\Omega_{0}$ given by equation (\ref{Oscfre}) can be used to determine the bent segment of the superfluid vortex lines, which in turn constrains the extent of the closed toroidal field lines region inside neutron stars. Therefore, our model may be used to quantify the magnetic field effects on the equation of state of neutron stars \citep{patra20}. Moreover, the damping time-scale of the oscillations provides the coupling properties of inner crust superfluid to the rest of the neutron star. This time-scale in turn can be used to put restrictions on the pairing energy close to the nuclear saturation density.

We applied time variable external torque and superfluid coupling model including vortex oscillations to the peculiar rotation evolution of PSR B0950+08.We used the data obtained and analysed by \citet{huang22} and employed our model to the spin-down rate of this pulsar for which data points with error bars are given in table 2 of their paper. We considered two cases for the form of the external torque. In Model I, we investigate the effects of a constant pulsar braking in short 14 yr observation interval given that external torque changes on much longer spin-down time-scale. In Model II, we examine the negative braking index measurement of this pulsar by invoking magnetic field change as the main cause for the time dependence of the external torque. The observed data is better explained with a temporary magnetic field growth on a time-scale $\tau_{\rm B}=B/\dot B\approx1.5\times10^{2}$~yr. In the course of secular evolution of spinning-down neutron stars, some of the core magnetic flux would be transported into the crust as vortices carries magnetic flux tubes with them \citep{srinivasan90}. This may be the underlying mechanism responsible for the seemingly magnetic field growth in old enough pulsars. Model fits to the data for PSR B0950+08 are shown in Fig. \ref{fig:fit}. Clustering of several spin-down rate data points around MJD 55000 may be as a result of short-circuit based magnetospheric noise \citep{cheng87} and its coupling with internal superfluid components \citep{erbil17}.  

Given that PSR B0950+08 has undergone only three distinct sinusoidal-like peaks in the spin-down rate and no successive iterative oscillations were observed in 40 yr of long-term data \citep{shaw22}, we can estimate $\tau_{\rm l}\sim20$ yr for the coupling time-scale of the linear regime crustal superfluid to the observed crustal normal matter. While obtaining the upper limit of 20 yr, we took advantage of the decrement in the amplitude of the latest increase in the spin-down rate mentioned in footnote 2, and  by applying the equation (\ref{model1}) we reached the conclusion.  If we use the transition value given by equation (\ref{lnlcreep}) for $kT/E_{\rm p}$ ratio and adopt the conservative value $\omega_{\rm cr}\approx4\times10^{-2}$ rad\,s$^{-1}$ \citep{zhou22}, then the non-linear creep recoupling time-scale $\tau_{\rm nl}=(kT/E_{\rm p})(\omega_{\rm cr}/|\dot\Omega_{\infty}|)$ becomes $\simeq1500$ yr. Since $\tau_{\rm l}\ll\tau_{\rm nl}$, this justifies our usage of linear creep approximation in our calculations.

Sinusoidal-like spin-down rate oscillations were also observed after the 1988 Christmas Vela glitch \citep{mcculloch90} and larger glitch in PSR B2334+61 \citep{yuan10}, indicating that reconfiguration of the vortex lines following glitches plays dominant role in torque oscillations acting on these neutron stars.

The long-term crustal magnetic field decay as a result of Ohmic diffusion and Hall drift processes may also cause alternating spin evolution in old pulsars \citep{zhang13}. Our approach here provides an alternative solution to the problem of anomalous neutron star spin evolution. According to our model, oscillations in the spin down rate of neutron stars could be either damped oscillations of the coupling between neutron star interior superfluid and normal matter crust, with extremely long damping time-scales extending to decades, excited by discrete events like glitches occuring before oscillation, or more likely resonant oscillation modes of the neutron star interior and crust driven by the magnetospheric oscillations of the external torque. For the former case, if not triggered by other events, the sinusoidal-like oscillations will decay on the time-scale given by equation (\ref{taulin}), and a relatively stable spin-down rate is again achieved. It seems that this first case represents the timing behaviour of PSR B0950+08. The latter may play a role in almost uniformly repeated oscillatory behaviour as seen from the cases of PSR B1540--06 and PSR B1828--11 \citep{shaw22}.

Our model equations can be applied to the observations of other sources showing oscillatory spin-down rate behaviour \citep{parthasarathy19,shaw22}. High cadence and long-term timing observations will enable us to probe into the magnetospheric braking mechanisms and superfluid properties of pulsars. A large sample of old isolated pulsars with upcoming high-precision timing, ultraviolet, or X-ray observations are essential ingredients that will help us for studying the evolution and interior characteristics of pulsars.

\section*{Acknowledgements}

This study was funded by Scientific Research Projects Coordination Unit of Istanbul University with project number MAB-2022-38210. We thank Professors Ali Alpar and Altan Baykal for very fruitful discussions. We are thankful to the referee for very constructive comments and suggestions which lead to significant improvement of presentation.

\section*{Data Availability}

No new data were generated in support of this theoretical study.








\bsp	
\label{lastpage}
\end{document}